\documentclass[twocolumn,amsmath,amssymb,prl]{revtex4-2}

\usepackage{graphicx}
\usepackage{dcolumn}
\usepackage{bm}
\usepackage{mathtools}
\usepackage{amssymb}
\usepackage{amsfonts}
\usepackage{amsmath}
\usepackage{float}
\usepackage{siunitx}
\usepackage{multirow}
\usepackage{hyperref}
\usepackage{xcolor}
\DeclareMathAlphabet{\mathpzc}{OT1}{pzc}{m}{it}

\begin{document}


\title{Magneto-Seebeck coefficient of Fermi-liquid  in three-dimensional Dirac/Weyl semimetal}
\author{F. R. Pratama$^{1}$}
\email{pratama.fr@aist.go.jp}
\author{Riichiro Saito$^{2}$}
\author{Nguyen T. Hung$^{2,3}$}
\email{nguyen@flex.phys.tohoku.ac.jp}

\affiliation{$^{1}$Mathematics for Advanced Materials-OIL, AIST, 2-1-1 Katahira, Aoba, 980-8577 Sendai, Japan\\ 
$^2$Department of Physics, Tohoku University, Sendai 980-8578, Japan\\
$^3$Frontier Research Institute for Interdisciplinary Sciences, Tohoku University, Sendai 980-8578, Japan}

\begin{abstract}
We investigate dissipationless magneto-Seebeck effect in three-dimensional Dirac/Weyl semimetal.  The Hall resistivity $\rho_{yx}$ and thermoelectric Hall coefficient $\alpha_{xy}$ exhibit plateaus at the quantum limit, where electrons occupy only the zeroth Landau level. In this condition, quantum oscillation in the Seebeck coefficient $S_{xx}\approx \rho_{yx}\alpha_{xy} $ is suppressed, and the massless fermions are transformed into a Fermi liquid system. We show that the Seebeck coefficient at the quantum limit is expressed by the harmonic sum of Fermi wavelength and  thermal de Broglie wavelength scaled by magnetic length.



\end{abstract}

\date{\today}
\maketitle


Research on thermoelectricity in low-dimensional materials is pioneered by the theoretical works of Hicks and Dresselhaus~\cite{hicks1993effect,hicks1993thermoelectric,heremans2013thermoelectrics}. They found that by reducing the dimension of materials (essentially confining charge carriers in one- or two-directions), larger Seebeck coefficient $S_{xx}$ (or thermopower) and power factor can be obtained due to enhancement of electronic density of states at the Fermi energy. Such quantum confinement effect takes place only if the typical size of the material is smaller than thermal de Broglie (TDB) wavelength~\cite{hung2016quantum,hung2018universal,hung2021origin}. In this Letter, we show that an analogous phenomenon also occurs in three-dimensional (3D) Dirac/Weyl semimetal when we apply magnetic field $B$. The confinement of massless fermions in a 3D Fermi pocket by $B$ will induce the Fermi liquid behavior in the material. 

An advantage of using a 3D Dirac/Weyl material in magneto-thermoelectric devices was pointed out by Skinner and Fu~\cite{skinner2018large}. They showed that a large, non-saturating $S_{xx}$ as a function of the magnetic field is achieved in the dissipationless limit. By keeping the number of carriers $n_0$ constant, the thermoelectric Hall coefficient over temperature $\alpha_{xy}/T$ is approaching a finite value that depends only on fundamental constants and Fermi velocity~\cite{kozii2019thermoelectric,han2020quantized,zhang2020observation}.  The Hall resistivity $\rho_{yx}$ is expressed by the classical formula $\rho_{yx}=B/(n_0e)$ ($e$ is the elementary charge), and as consequence the $S_{xx}$ is linearly proportional to $B$, $S_{xx} \approx \rho_{yx}\alpha_{xy} \propto B$~\cite{generalization}. However, Galeski~\textit{et al.}~\cite{galeski2021origin} recently observed saturating  $S_{xx}$ in a nearly massless~\cite{chen2015magnetoinfrared}, 3D Dirac semimetal ZrTe$_5$ in magnetic field $B\approx 2 ~\mathrm{T}$, in which 3D quantum Hall effect (QHE) occurs. They showed that the plateau of the Hall resistivity is proportional to the half of the Fermi wavelength in the direction parallel to the magnetic field, $\lambda_{F}^{\parallel}$.

Similar experimental results are reported by Tang et al.~\cite{tang2019three} in ZrTe$_5$, and by Wang et al.~\cite{wang2020approaching} in HfTe$_5$. It is important to note that the QHE appears only in two-dimensional (2D) electron systems when the Fermi level is located between two Landau levels (LLs)~\cite{stormer1986quantization,halperin1987possible}. Thus, 3D materials normally do not exhibit the QHE because of the continuous dispersion along the direction of $B$. To explain the QHE in the 3D Dirac/Weyl semimetal, it was argued~\cite{tang2019three} that the formation of charge density wave (CDW)~\cite{sehayek2020charge,qin2020theory} due to the Fermi surface instability plays a decisive role in discrete LL near the Fermi energy. In other words, the CDW transforms the 3D system into a stack of decoupled 2D layers, and each layer conducts the Hall current on the edges. However, spectroscopic and transport measurements ~\cite{galeski2021origin} do not show any signs of CDW. Thus, it is necessary to confirm analytically whether the Hall plateau arises from the intrinsic bulk properties of the 3D Dirac/Weyl fermion, especially since recent calculations on the magneto-transport coefficients of the Weyl semimetal with finite thickness~\cite{chang2021three,chang2021threeb,ma2021three,xiong2022understanding} do not show such feature that $\rho_{yx}\propto\lambda_{F}^{\parallel}/2$.

In this Letter, we show that the saturated value of $S_{xx}$ becomes the Seebeck coefficient of 3D Fermi liquid~$\mathcal{S}$~\cite{behnia2016nernst}. In particular, we show that the condition for $S_{xx} \equiv \mathcal{S}$ follows a scaling law between three fundamental lengths: the magnetic length, the Fermi wavelength, and the TDB wavelength in the direction perpendicular to $B$, where charge carriers are magnetically confined. Further, we discuss that the condition originates from the existence of the zeroth LL in the 3D Dirac/Weyl fermion, which is independent of $B$, and thus can not be satisfied in electron gas in conventional 3D metals or semiconductors. 

The calculation of the Hall conductivity and thermoelectric Hall coefficient is carried out within quantum edge formalism~\cite{halperin1982quantized,girvin1982inversion,bergman2010theory,skinner2018large,kozii2019thermoelectric}, where transport scattering time $\tau$ is assumed to be small compared with the inverse of cyclotron frequency $1/\omega_c$, i.e. $\omega_c\tau \gg 1$. In this condition, the magnitude of the longitudinal conductivity $\sigma_{xx}$ is much smaller than $\sigma_{xy}$, which implies that the Hall resistivity is approximately given by $\rho_{yx}\approx 1/\sigma_{xy}$, and $S_{xx}\approx \rho_{yx}\alpha_{xy}$~\cite{von1986quantized}. Let us consider a 3D Weyl/Dirac semimetal with a volume $L_x L_y L_z $ , where $L_i$ with $i = x, y, z$ is the length of the material in the $i$-th direction. In the presence of an external magnetic field in the $z$-direction $\mathbf{B} = B\hat{\mathbf{z}}$, an electrons is confined to move on an orbit in the $xy$-plane, but move freely in the $z$-direction with the Fermi velocity $v_{F}^{z}$. A potential difference $V_x$ is applied between the right ($x=+L_x/2$) and left ($x=-L_x/2$) edges of the sample, which gives rise to the electric field $\mathbf{E}=E_x\hat{\mathbf{x}}=-\partial V_x / \partial x$. In the presence of both $B$ and $E_x$, the Landau level (LL) spectrum without the Zeeman term is given by (see Sec. A in the Supplemental Material for the derivation):
\begin{align}
    E_{n}(B,k_y,k_z) = \varepsilon_d(B,k_y) +\varepsilon_n(B,k_z),
    \label{eq:LL}
\end{align}
where $\varepsilon_d(k_y)\equiv \hbar k_y v_d$ and $\varepsilon_n(k_z)\equiv\mathrm{sgn}(n) \sqrt{{\mathcal{E}_B}^2|n|/\gamma^3 + \left(\hbar v_{F}^{z}  k_z/\gamma\right)^2}$. Because of $\varepsilon_d(k_y)$, the LL is dispersive along $k_y$, which allows charge carriers to move in the $y$-direction with velocity $v_y = \hbar^{-1}\partial \epsilon_n(k_y,k_z)/\partial k_y = v_d$, where $v_d\equiv E_x/B$ is the drift velocity. $\varepsilon_n(k_z)$ corresponds to the cyclotron energy, where $\mathcal{E}_B \equiv \sqrt{2}\hbar v_{F}^{\perp}/\ell_B$, $\ell_B\equiv\sqrt{\hbar/(e B)}$ is the magnetic length, and $v_{F}^{\perp}$ is the Fermi velocity in the direction perpendicular to the magnetic field. Due to the relativistic nature of the Dirac fermion, the effect of $E_x$ is also included in the Lorentz factor, $\gamma \equiv 1/\sqrt{1-(v_d/v_F^{\perp})^2}$, which induces the shrinking of the LL for $E_x\neq 0$~\cite{peres2007algebraic,arjona2017collapse}. Nevertheless, since we focus on the case for strong $B$, we can take $\gamma\approx 1$. It is noted that for $|k_z|\neq 0$, there are two possible values of the zeroth LL, i.e.  $E_{\pm 0}=\varepsilon_d(k_y)\pm \hbar v_{F}^{z} k_z$. 

By using the fact that $\ell_B\ll L_z$, $k_z=2\pi n_z/L_z$ ($n_z\in \mathbb{Z}$) can be treated as a continuous variable $\sum_{k_z}/L_z\rightarrow \int_0^{\infty}dk_z/\pi$, the density of states per unit volume (DOS) for each spin direction is given by 
\begin{align}
    \mathcal{D}(E) = \frac{eB}{h}\sum_{k_y}\int\limits_{0}^{\infty}\frac{dk_z}{\pi}\sum_{n=-\infty}^{\infty}N_n\delta\left[E - E_{n}(k_y, k_z)\right],
    \label{eq:dos0}
\end{align}
where we define $N_n\equiv 1/2~(1)$ for $n=0~(n\neq 0)$ to avoid double counting of the zeroth LL that includes both $E_{+ 0}$ and $E_{- 0}$.
The number of charge carriers $n_0$ is given by the difference between the number of electrons and holes, as follows:
\begin{align}
   n_0=\int\limits_{0}^{\infty}\mathrm{d}E\mathcal{D}(E)f(E-\mu^{*}) - \int\limits_{-\infty}^{0}\mathrm{d}E\mathcal{D}(E)[1-f(E-\mu^{*})],
   \label{eq:n0}
\end{align}
where $f(E-\mu^*)\equiv 1/[e^{\beta(E-\mu^*)}+1]$ is the Fermi-Dirac distribution function, $\beta\equiv 1/(k_BT)$. The chemical potential $\mu^{*}$ is given as a function of $x$ due to $V_x$:
\begin{align}
  \mu^{*}(x)=\mu + {eV_x x}/{L_x},
  \label{eq:mu}
\end{align}
where $\mu$ is a constant which determines the average carrier density~\cite{gerhardts2013} in the sample. Hereafter, we adopt the condition that $\mu$ in Eq.~\eqref{eq:mu} is kept constant with respect to magnetic field. 


For a given $n_0$ in Eq.~\eqref{eq:n0}, the electrical $J_y$ and heat $J_y^Q$ currents are given by~\cite{kozii2019thermoelectric}
\begin{align}
    \binom{J_y}{J_{y}^{Q}} = v_yn_0\binom{-e}{E-\mu^{*}} =  -E_x\binom{\sigma_{xy}}{T \alpha_{xy}}.
    \label{eq:Jyb}
\end{align}

By considering only the currents along the edges $(x=\pm L_x/2)$, $\sigma_{xy}$ and $\alpha_{xy}$ in Eq.~\eqref{eq:Jyb} are given as a function of $\mu$ and $T$, as follows (see Secs. C and D in the Supplemental Material for the derivation):
\begin{align}
    \sigma_{xy} &= \frac{2e^2}{h}\sum\limits_{j=\pm 1}\sum\limits_{n=0}^{\infty}j{N}_n   \int\limits_{0}^{\infty}\frac{dk_z}{\pi}f\left[\varepsilon_{n}(B,k_z) -j\mu\right]\nonumber\\
    & \equiv \sigma_{xy}^{(0)} + \sum\limits_{n=1}^{\infty}\sigma_{xy}^{(n)},
    \label{eq:sxy4}
\end{align}
and
\begin{widetext}
\begin{align}
    \alpha_{xy} = \frac{2e k_B}{h} \sum\limits_{j=\pm 1}\sum\limits_{n=0}^{\infty}{N}_n \int\limits_{0}^{\infty} \frac{dk_z}{\pi}
    \left\{  \ln\left[1+e^{-\beta\left\{\varepsilon_n(B,k_z) - j\mu\right\}}\right]  
    + \frac{\beta\left[\varepsilon_n(B,k_z)-j\mu\right]}{1+e^{\beta\left[\varepsilon_n(B,k_z)-j\mu\right]}} \right\}\equiv \alpha_{xy}^{(0)} + \sum\limits_{n=1}^{\infty}\alpha_{xy}^{(n)},
    \label{eq:axy}
\end{align}
\end{widetext}
where $\sigma_{xy}^{(n)}$ and $\alpha_{xy}^{(n)}$ denote, respectively, the Hall conductivity and thermoelectric Hall coefficient for $n$-th LL. The prefactor of 2 in Eqs. (\ref{eq:sxy4}) and (\ref{eq:axy}) is given by assuming that filling factor of the LL at $x=\pm L_x/2$ are the same~\cite{cage1997,panos2014}, where almost symmetrical currents at the both edges flow in the same direction~\cite{gerhardts2013,haremski2020}. The index $j= +1$ and $-1$ indicate the contributions of electron and hole, respectively. $\sigma_{xy}^{(0)}$ and $\alpha_{xy}^{(0)}$ are the Hall conductivity and thermoelectric Hall coefficient for the zeroth LL, which are obtained analytically as follows:
\begin{align}
    \sigma_{xy}^{(0)} =\frac{2 e^2 }{ h^2 v_{F}^z \beta }\sum\limits_{j=\pm 1} j\mathrm{ln}\left( 1+e^{j\beta\mu} \right),
    \label{eq:sxy00}
\end{align}
and
\begin{equation}
\begin{split}
    \alpha_{xy}^{(0)}  =  \frac{-2 e k_B }{ h^2 v_{F}^z}  \sum\limits_{j=\pm 1} \Bigg[ \frac{2}{\beta}\mathrm{Li}_2(-e^{j\beta\mu})+j\mu\ln{(1+e^{j\beta\mu})}\Bigg],
    \label{eq:axy00}
\end{split}
\end{equation}
where $\mathrm{Li}_2(z)\equiv\sum_{k=1}^{\infty}z^k/k^2$ is the dilogarithm function. 

Hereafter, we adopt the parameters for calculating $\rho_{yx}=1/\sigma_{xy}$ and $\alpha_{xy}$ from the experimentally determined values of 3D Dirac material ZrTe$_{5}$~\cite{galeski2021origin}. The material has an orthorombic crystal structure and anisotropic Fermi velocities $v_{F}^{i}$ along $i=a,~b,$ and $c$-axis, where $v_{F}^{a}=1.164\times 10^5$ m/s, $v_{F}^{b}=1.534\times 10^4$ m/s, and $v_{F}^{c}=3.489\times 10^5$ m/s. By aligning $B$ parallel to the $b$-axis, the Fermi wavevector $k_{F}^{b}= 72.9\times 10^{-3}\mathrm{\AA}^{-1}$ is extracted from the Hall plateau in the quantum limit, where only the zeroth LL is occupied. In order to reproduce experimental measurement of $\rho_{yx}$ with our model, we use $k_{F}^{z} = k_{F}^{b}$, $v_{F}^{z} = v_{F}^{b}$, $v_{F}^{\perp} = \sqrt{v_{F}^{a}v_{F}^{c}}=2.015\times 10^5$ m/s, and $\mu = \hbar v_{F}^{z} k_{F}^{z} = 7.36$ meV.


\begin{figure}[t]
\begin{center}
\includegraphics[width=65mm]{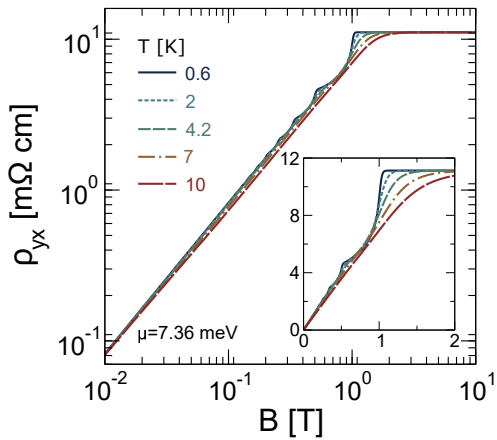}
\caption{Log-log plot of $\rho_{yx}$ as a function of $B$ for a constant $\mu=7.36~\mathrm{meV}$, for several values of $T$. Inset shows the linear plot of $\rho_{yx}$ with $B=0-2~\mathrm{T}$. }
\label{fig:hall}
\end{center}
\end{figure}


In Fig.~\ref{fig:hall}, we show a log-log plot of  $\rho_{yx}$ as a function of $B$ for $T= 0.6, 2, 4.2, 7$ and $10~\mathrm{K}$. For all temperatures, the quantum Hall regime occurs at $B\sim~1~T$, as shown in the inset of Fig.~\ref{fig:hall}. It is noted that the quantization of $\rho_{yx}$ is not as pronounced as in the two-dimensional case, due to the logarithmic singularity of the DOS upon the integration on $k_z$ (see Sec. B in the Supplemental Material).  As we further increase $B$, $\rho_{yx}$ becomes independent of $B$ and $T$, which indicates that all $n> 0$ LLs are depopulated. As shown in the inset of Fig.~\ref{fig:hall}, the Hall plateau at $\rho_{xy} \approx 11.6~\mathrm{m}\Omega\cdot\mathrm{cm}$ occurs for $B \geq 1.2~\mathrm{T}$. The present calculation reproduces recent experiments by Tang\cite{tang2019three} and Galseki~\cite{galeski2021origin}. The good agreement can be attributed to small effective mass in $\mathrm{ZrTe_5}$~\cite{galeski2021origin,jiang2017landau,tang2019three,zheng2016transport,wang2018vanishing} (in the order of $\sim 0.01m_e$, where $m_e$ is the mass of free electron), and because there is only one spin degeneracy for each zeroth LL~\cite{chen2015magnetoinfrared,zhang2019anomalous,chen2017spectroscopic}. 

\begin{figure}[t]
\begin{center}
\includegraphics[width=65mm]{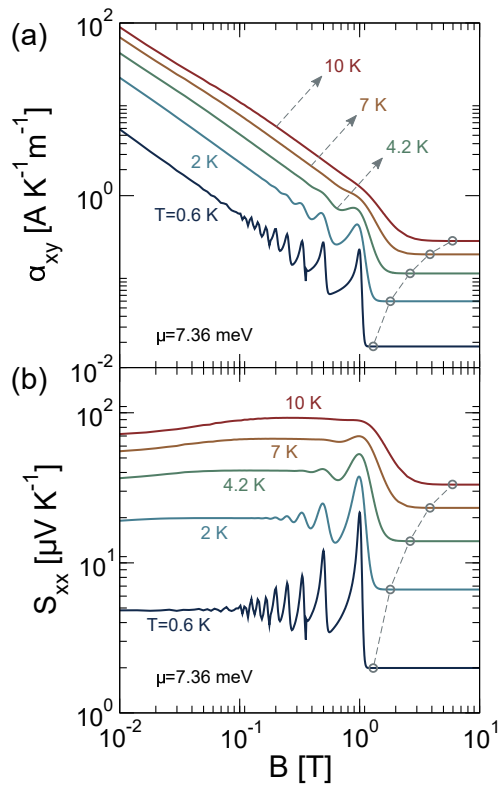}
\caption{ Log-log plot of (a) $\alpha_{xy}$ and (b) $S_{xx}$ as a function of $B$ for several values of $T$. The chemical potential is set by $\mu=7.36~\mathrm{meV}$. Grey circles indicates the field $B=\mathcal{B}(\mu,T)$ where $\alpha_{xy}$ and $S_{xx}$ begin to saturate. }
\label{fig:peltseeb}
\end{center}
\end{figure}

Let us discuss the quantum limit case, in which $\rho_{yx}(\mu,T) = 1/ \sigma_{xy}^{(0)} (\mu,T)$. From Eq.~(\ref{eq:sxy00}), the term at $j=-1$ vanishes for low temperature ($|\mu\beta|\gg 1$), which means that only electrons contribute to the constant $\rho_{yx}$ for $\mu>0$ (and holes for $\mu<0$ but with opposite sign of $\rho_{yx}$). By using $\mathrm{ln}(1+e^{\beta\mu})\approx \beta\mu$, $\rho_{yx}$ is independent of $T$ as follows:
\begin{align}
    \rho_{xy} = \frac{2e^2\mu}{h^2 v_{F}^{z}}=\frac{\pi h}{e^2 k_{F}^{z}} = \frac{h\lambda_{F}^{z}}{2 e^2},
    \label{eq:hplateau}
\end{align}
where $\lambda_{F}^{z} = 2\pi /k_{F}^{z} \equiv \lambda_{F}^{\parallel}$. 
It is noted that the magnitude of the Hall plateau shown by Eq.~(\ref{eq:hplateau}) has been confirmed experimentally in the samples of $\mathrm{ZrTe_5}$ with thickness $L_z\approx 100~\mathrm{\mu m}$, while an experiment on the 2D Dirac semimetal $\mathrm{Cd_3As_2}$~\cite{zhang2019quantum} with $L_z\approx 60-70~\mathrm{nm}$ does not indicate such behaviour. Thus, we show that Eq.~(\ref{eq:hplateau}) is indeed an intrinsic property of 3D Dirac semimetal.

In Figs.~\ref{fig:peltseeb} (a) and (b), we show log-log plots of $\alpha_{xy}$ and $S_{xx} = \alpha_{xy}/\sigma_{xy}=\rho_{yx}\alpha_{xy}$ as a function of $B$ for $T= 0.6, 2, 4.2, 7$ and $10$ K, respectively. We can see that for $B\leqslant 1~\mathrm{T}$, $\alpha_{xy}$ decreases linearly with increasing $B$, which means that $\alpha_{xy} $ is inversely proportional to $ B$ in the linear scale. On the other hand, $S_{xx}$ does not show significant dependence on $B$. As $B$ becomes larger, oscillations appears in both $\alpha_{xy}$ and $S_{xx}$ for $T\leqslant 4.2~\mathrm{K}$ since a few LLs touch the chemical potential. Due to the smearing of the Fermi-Dirac distribution function by increasing $T$, the individual peaks in the oscillation become indistinguishable. As the quantum limit takes place, $\alpha_{xy}$ and $S_{xx}$ become constants for all temperatures. By applying identity $\mathrm{Li}_2(z) + \mathrm{Li}_2(1/z) = -\pi^2/6 - [\mathrm{ln}(-z)]^2/2$ on Eq.~(\ref{eq:axy00}), the explicit formula for thermoelectric Hall plateau $\alpha_{xy} = \alpha_{xy}^{(0)}$ is given as follows:
\begin{align}
    \alpha_{xy} = \frac{2\pi^2}{3}\frac{{e k_B}^2}{h^2 v_{F}^z} T.
    \label{eq:thplateau}
\end{align}
Eq.~(\ref{eq:thplateau}) was first obtained by Kozii et al.~\cite{kozii2019thermoelectric} as the saturating limit for $\alpha_{xy}$ of an ideal 3D Dirac material at a large $B$, with a fixed number of carriers. We complement the result by showing that exactly the same formula also prevails for a fixed chemical potential. By combining Eqs.~(\ref{eq:hplateau}) and (\ref{eq:thplateau}), $S_{xx}$ at the quantum limit is given by
\begin{align}
    S_{xx} =\frac{\pi^2}{3}\frac{{k_B}^2}{e}\frac{T}{\mu} \equiv \mathcal{S},
    \label{eq:splateau}
\end{align}
where $\mathcal{S}$ is the expression for the Seebeck coefficient of the Fermi liquid system~\cite{behnia2016nernst}.


\begin{figure}[t]
\begin{center}
\includegraphics[width=75mm]{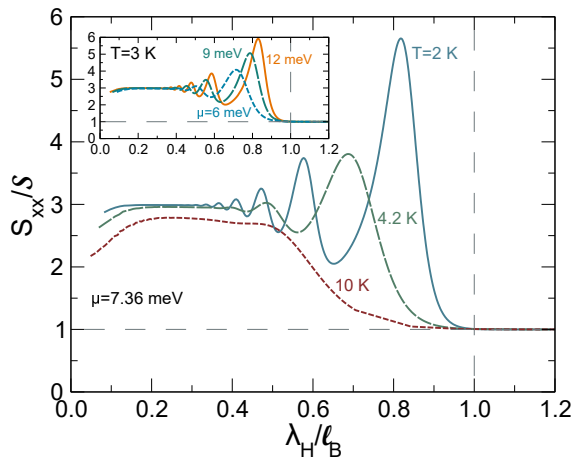}
\caption{ Plot of the scaled Seebeck coefficient $S_{xx}/\mathcal{S}$ as a function of $\lambda_H/\ell_B$ for constant $\mu=7.36~\mathrm{meV}$ for several values of $T$ (straight lines). Inset: $S_{xx}/\mathcal{S}$ at a constant $T=3~\mathrm{K}$ for several values of $\mu$. }
\label{fig:ferm}
\end{center}
\end{figure}

Now, let us discuss the onset of magnetic field, $\mathcal{B}$, in which the 3D Dirac/Weyl material behaves as the Fermi liquid system at $\mathcal{B}(\mu,T)$ (open circles in Figs.~\ref{fig:peltseeb}). It is noted that in the quantum limit, the uncertainty principle determines that the radius of electron's orbit is given by the magnetic length $\ell_B$. Therefore, it is useful to express the dependence of $\mathcal{B}$ in term of the length scales associated with $\mu$ and $T$, in the direction perpendicular to $B$ where magnetic confinement of electrons occurs. At $T=0~\mathrm{K}$, the plateau is formed when the magnetic energy is larger than the chemical potential, i.e. $\mathcal{E}_B \geqslant \mu$ with $\mu=\hbar v_{F}^{\perp} k_{F}^{\perp} = h v_{F}^{\perp}/ \lambda_{F}^{\perp} $. Thus, near the absolute zero, $S_{xx}$ becomes $\mathcal{S}$ when the ratio of the two length scales satisfies $\ell_B/\lambda_{F}^{\perp}  \leqslant 1/(\sqrt{2}\pi) \approx 0.225$.

For given a finite $T$, $\sigma_{xy}$ and $\alpha_{xy}$ are obtained by a convolution of $-\partial f(E-\mu,T)/\partial E = (\beta/4)\mathrm{sech}^2[\beta(E-\mu)/2] $ with those quantities at $T=0~\mathrm{K}$ (see Supplemental material). Thus, the shift from $\ell_B/\lambda_{F}^{\perp}$ is given by $W = 2k_B T/\mathcal{E}_B$, resulting in Eq.~(\ref{eq:splateau}) holds if following condition is satisfied:
\begin{align}
    \frac{\ell_B}{\lambda_{F}^{\perp}} + W \leqslant \frac{1}{\sqrt{2}\pi}.
    \label{eq:scale1}
\end{align}
The length scale associated with the temperature is thermal de Broglie wavelength~\cite{hung2021origin}. We adopt a definition of the thermal de Broglie wavelength~\cite{yan2000general} which is suitable for massless particle $\Lambda^{\perp}\equiv h v_{F}^{\perp}/(2\pi^{1/3}k_B T)$.  
Equivalently, $W = \sqrt{2}\pi^{2/3}\ell_B/\Lambda^{\perp}$. By defining $1/\lambda_{H}$ as a weighted harmonic sum of $\lambda_{F}^{\perp}$ and $\Lambda^{\perp}$:
\begin{align}
   \frac{1}{\lambda_H} = \frac{\sqrt{2}\pi}{\lambda_{F}^{\perp}} + \frac{2\pi^{5/3}}{\Lambda^{\perp}},
   \label{fig:harmonic}
\end{align}
the condition for obtaining $S_{xx} = \mathcal{S}$ is given by
${\lambda_{H} }/{\ell_B}\geqslant 1$.
In Fig~\ref{fig:ferm}, we plot $S_{xx}/\mathcal{S}$ with $\mu=7.36~\mathrm{meV}$ for $T=2,~4.2$, and $10~\mathrm{K}$. Here, we can see the suppression of quantum oscillation for ${\lambda_{H} }/{\ell_B}\geqslant 1$. In Fig~\ref{fig:ferm}, $\Lambda^{\perp}$ is varying while $\lambda_{F}^{\perp}$ remains constant. For the opposite condition, we plot $S_{xx}/S$ for $\mu=6,~9$, and $12~\mathrm{K}$ at a constant $T=3~\mathrm{K}$ (see inset plot in Fig~\ref{fig:ferm}), where similar behaviour is observed. This phenomenon shows that the confinement within the length of $\ell_B$ will transform the 3D Dirac fermions into a Fermi liquid system. In other words, the states of the Dirac fermions at the zeroth LL become localized in the $k$-space within a small Fermi surface (due to low carrier density $n_0$), which does not change with keeping chemical potential constant. In contrast, for an ordinary massive 3D electron gas with a fixed $\mu$, we show in the Section E of Supplemental material that $\rho_{yx}$ tends to increase after $n=1$ LL is emptied, while $\alpha_{xy}$ becomes zero. This phenomenon is unlikely to occur because of large $n_0$, which tends to be preserved in order to minimize the change of energy~\cite{galeski2021origin}.

Finally, we have to consider the ranges of $B$ and $T$ in which the Fermi liquid phase can survive. For $\lambda_H/\ell_B\leqslant 1.2$ in Fig.~\ref{fig:ferm}, the maximum value of $B$ is around $6~\mathrm{T}$ for $T\leqslant 10~\mathrm{K}$ and $\mu \leqslant 12~\mathrm{meV}$. Around this  $B$ and higher, the scaling law may be broken if the Zeeman term becomes more noticeable, especially if the material has a large $g$-factors. This may explain the increase of $\rho_{xy}$ in $\mathrm{ZrTe_5}$~\cite{galeski2021origin,tang2019three} and $\mathrm{HfTe_5}$~\cite{wang2020approaching} for $B\geqslant 3$ T though the Hall plateaus are observed for $B<3$ T. In these materials, the value of $g$ is in the order $\sim 10$~\cite{galeski2021origin,chen2017spectroscopic,jiang2017landau,liu2021induced}). It also has been observed that large $B$ can drive phase transitions such as band inversion~\cite{zhang2019anomalous,zheng2017field}, anomalous QHE~\cite{liu2021induced}, as well as  fractional QHE~\cite{tang2019three} and mass generation~\cite{liu2016zeeman} due to many body interactions. Also at higher temperature, the transport becomes dissipative due to scattering effects~\cite{han2020quantized}, where $\rho_{xx}$ increases, and as a consequence, the relation $S_{xx}\approx \rho_{yx}\alpha_{xy}$ is no longer valid. Nevertheless, our proposed scaling law for an ideal 3d Dirac material may be observed experimentally within proper ranges of $B$ and $T$.

In summary, we have analytically shown that at a fixed chemical potential, both the Hall resistivity and thermoelectric Hall coefficient show plateau structures. In this condition, magnetic confinement of carriers occupying the zeroth LL transform the 3D Dirac/Weyl fermion into the Fermi liquid. Particularly, the threshold of this phenomenon can be parameterized by magnetic length 
, the Fermi wavelength, and the thermal de Broglie wavelength. Our ﬁndings
establish a relationship between the three fundamental lengths in the quantum Hall and thermoelectric Hall effects in the 3D Dirac/Weyl semimetals.

\section*{Acknowledgments}
N.T.H. acknowledges JSPS KAKENHI (Grant No. JP20K15178) and the financial support from the Frontier Research Institute for Interdisciplinary Sciences, Tohoku University. R.S. acknowledges JSPS KAKENHI Grant No JP18H01810 and CSIS, Tohoku University.


%

\end{document}